# Relation between stochastic processes and thermodynamics of trajectories

**V. V. Ryazanov**

Institute for Nuclear Research, pr. Nauki, 47 Kiev, Ukraine, e-mail: vryazan19@gmail.com

Highlights

-A correspondence was obtained between the roots of the Lundberg equation and the description of the first passage time.
-Risk theory is used in the description of dynamic activity.
-The parameter conjugate to a random variable is considered as an external field.
-A combination of approaches from statistical physics and the theory of random processes is proposed.

The process of fluctuations of trajectory observables of stochastic systems is related to processes with independent increments from the risk theory. The first-passage times of variables of the thermodynamics of trajectories, in particular, dynamic activity, are considered. A correspondence between expressions of the theory of random processes and thermodynamics of trajectories, as well as deviations from such a correspondence for the process of fluctuations of trajectory observables, are established. The connections between general regularities of first-passage times in the theory of random processes and thermodynamics of trajectories are discussed. A more complete use of the theory of random processes in physical problems is proposed, and the possibilities of combining approaches of the theory of random processes and statistical physics are indicated.

## 1. Introduction

The theory of random processes and statistical physics are close to each other in the nature of their methods and the problems they study. Statistical physics uses many results of the theory of random processes. Both sciences developed independently of each other. The theory of random processes uses mathematical methods, and statistical physics uses experimental results (although stochastic experiments are also considered in probability theory). Each of the sciences has established independent methods and approaches. It may be useful to bring these methods closer together. These kinds of problems are considered in this article using physical examples.

As an example of the connection between statistical physics and the theory of random processes, we can cite randomly stopped processes, Ref. [1] $\xi(\theta_s)$, which contain an exponentially distributed random variable $\theta_s$, independent of the process $\xi(t)$, with the characteristic function (3). Averaging is carried out with an exponential distribution over time of the characteristic function of the process $\xi(t)$. The same procedure, as shown in Ref. [2], is carried out in the method of nonequilibrium statistical operator Refs. [3-4], corresponding to the characteristic function of the process $\xi(t)$ (2). As another example, we point to the consideration of a two-level system in Refs. [5] and [6]. The results of considering this system, carried out in Ref. [5] using lattice Poisson processes, correspond to the results obtained in Ref. [6] using thermodynamics of trajectories. However, as will be shown below, this is not always the case. For example, thermodynamics of trajectories considers only positive quantities, while the theory of random processes also considers negative ones.

First-passage time (FPT) and other boundary functionals of random processes are widely and effectively used to study various nonequilibrium phenomena (Refs. [5, 7-11]). FPT of a stochastic process across a boundary is important for many important and still developing applications. There are various approaches to studying such problems. Many of them are based on the theory of random processes. The theory of random processes [12-22], in particular, risk theory [18-22], is compared in this paper with the thermodynamics of trajectories. The thermodynamics of trajectories, Refs. [10-11, 23-34] is a new and promising direction in nonequilibrium thermodynamics.

The use of risk theory results allows one to explicitly write expressions for the first-passage time. In Refs. [35–37], expressions for limiting cumulants are written out, which are of a universal nature. For these cumulants, valid for many systems and situations, boundary functionals are defined, primarily the first-passage time. A comparison of the calculated values obtained for such models is given in Refs. [37-38]. Explicit connections between the mathematical theory of random processes and the results of nonequilibrium thermodynamics based on experimental data can be useful both for these areas and in various applications. In addition to explicit results for thermodynamics of trajectories obtained by probabilistic methods, a comparison of approaches and methods of thermodynamics of trajectories and the theory of random processes is also of interest. Such a comparison is given in Sections 3 and 4. Many probability relations, after physical interpretation, can be useful in physical research, providing new, more general possibilities. However, the results of statistical physics and nonequilibrium thermodynamics can also be useful for the theory of random processes.

The connections between the theory of random processes and statistical physics are very extensive. In statistical physics, renewal theory (e.g., Ref. [39]), martingales Refs. [40–41], various random processes, FPT, and other results of the theory of random processes are widely used. In this paper, the theory of random processes and the thermodynamics of trajectories are compared. Only some aspects of such a comparison are considered. It should be noted that probabilistic methods are used in the thermodynamics of trajectories. For example, the initial equation is the stochastic master equation for the process probabilities. The function θ(s) (7) is then given by the largest eigenvalue of the tilted master operator. In this paper, inquiries that have not been previously addressed in the thermodynamics of trajectories are presented.

The article is organized as follows: Section 2 provides an overview of the problems discussed. Section 3 compares risk theory and the thermodynamics of trajectories. General relationships between the theory of random processes and the thermodynamics of trajectories are considered. Section 4 provides estimates of dynamic activity obtained from the relationship between risk theory and thermodynamics of trajectories. Section 5 discusses the results obtained.

## 2. Review of the issues under consideration

### 2. 1. Review of the used results of the theory of random processes

Statistical physics and thermodynamics use a number of definitions from probability theory. Firstly, we point out the well-known fact that the statistical sum Z of the canonical distribution is the Laplace transform of the density of the distribution of microstates by energies; ω(E) is the probability density function of states per unit energy range in the partition function $Z_{can} = \int dE\,\omega(E)e^{-\beta E}$ .

Let $X$ be a random variable with cumulative distribution function $F_X$. The moment generating function (*MGF*) of $X$ (or $F_X$), denoted by $M_X(t)$, is:

$$M_X(t)=\boldsymbol{E}[e^{tX}] \tag{1}$$

provided this expectation exists for $t$ in some open neighborhood of 0. That is, there is an $h>0$ such that for all $t$ in $-h<t<h$, $\boldsymbol{E}[e^{tX}]$ exists. If the expectation does not exist in an open neighborhood of 0, we say that the moment generating function does not exist.

The *MGF* is the two-sided Laplace transform of the distribution of a random variable (up to reflection). For non-negative random variables, the characteristic function is a special case of the Laplace-Stieltjes transform. Thus, the mathematical definition of the *MGF* (and the characteristic function and Laplace transform) correspond to or are exactly equal to the definition of the partition function in statistical physics.

A process with independent increments $\xi(t)$ is called homogeneous if it is defined on $[0,\infty)$, and the distribution $\xi(t+h)-\xi(t)$ coincides with the distribution $\xi(h)$ for all $t>0$ and $h>0$. The characteristic function (*CF*) of a homogeneous process $\xi(t),\ t\geq 0$ is determined in the theory of random processes (for $\xi(0)=0$) (Refs. [12-16]) by the relation:

$$\boldsymbol{E}e^{i\alpha\xi(t)}:=\int_{-\infty}^{\infty}e^{i\alpha x}dF(x)=e^{t\Psi(\alpha)},\ t\geq 0, \tag{2}$$

where $F(x)=P(\xi<x)$ is the distribution function of a random process $\xi(t),\ t\geq 0$, and the function $\psi(\alpha)$ represents the scaled cumulant generating function (*SCGF*) (cumulant) of the process $\xi(t),\ t\geq 0$. The *SCGF* does not depend on $t$ and characterizes all finite-dimensional distributions of the process.

For the characteristic function of the process $\xi(\theta_s)$, defined as:

$$\boldsymbol{E}e^{i\alpha\xi(\theta_s)}=s\int_{-\infty}^{\infty}\boldsymbol{E}e^{i\alpha\xi(t)}e^{-st}dt, \tag{3}$$

we consider $\xi(\theta_s)$ as a randomly stopped process, Ref. [1]. Randomly stopped processes include an exponentially distributed random variable $\theta_s$ that is independent of the process $\xi(t)$, $P\{\theta_s>t\}=e^{-st},\ s>0, t>0$. The characteristic function of a randomly stopped process is the Laplace-Carson transform of the characteristic function given by equation (2).

Some boundary functionals considered below describe physical quantities characterizing time. These quantities are positive, and their characteristic functions transform into the Laplace transform, which have a probabilistic meaning. For example, for the Laplace transform $\boldsymbol{E}e^{-\gamma\tau^+}=\int_0^{\infty}e^{-\gamma\tau^+}f(\tau^+)d\tau^+$ of the distribution density function of the quantity $\tau^+$, the moment of the first exit for the level $x>0$, $\tau^+(u)=\inf\{t:\xi(t)>u\}$, the equality $\boldsymbol{E}e^{-\gamma\tau^+}=P\{\tau^+\leq\tau\}$, $P\{\tau>t\}=e^{-\gamma t}$, is true, where $\tau$ is a random variable independent of $\xi=\tau^+$, and having an exponential distribution with parameter $\gamma$.

For the characteristic function (3) of $\xi(\theta_s)$ the following relation is satisfied:

$$\varphi(s,\alpha):=\boldsymbol{E}e^{i\alpha\xi(\theta_s)}=\frac{s}{s-\Psi(\alpha)}, \tag{4}$$

and also, the fundamental factorization identity, which is also the identity of infinitely divisible factorization:

$$\varphi(s,\alpha)=\varphi_+(s,\alpha)\varphi_-(s,\alpha),\qquad \varphi_\pm(s,\alpha)=\boldsymbol{E}e^{i\alpha\xi^\pm(\theta_s)}, \tag{5}$$

where $\xi^\pm(t)=\sup_{0\leq t'\leq t}(\inf)\xi(t')$ represents the extremes of the process $\xi(t')$ on the interval $[0, t]$.

If for a process $\xi(t)$, $t \geq 0$, the function $\psi(\alpha)$ from (2) at $i\alpha = r$ is equal to *s*, then we obtain the equation:

$$\Psi(\alpha)\big|_{i\alpha=r} := k(r) = s, \quad \pm \operatorname{Re} r \geq 0\,. \tag{6}$$

This equation, in risk theory (Refs. [18-22]) is referred to as the fundamental Lundberg equation.

SCGF (2) in the thermodynamics of trajectories is interpreted as an analog of nonequilibrium free energy, Ref. [24]. However, Lundberg equation (6) has an ambiguous interpretation. If k(r) is free energy [24], then s in equilibrium has the meaning U-TS where U is the internal energy, S is the entropy, T is the temperature. But in equation (6) there is a nonequilibrium situation. In Ref. [41] a similar relation is used to determine the number of roots of this equation and to find them.

Due to the convexity $k(r)$ in the neighborhood of zero ($k``(0) > 0$) for sufficiently small *s* for continuous from above (from below) $\xi(t)$, $t \geq 0$, equation (6) has a positive root $r_s = \rho_+(s) > 0$ (negative root $r_s = -\rho_-(s) < 0$). The behavior of the roots $\rho_\pm(s)$ at $s \to 0$ depends on the mean value $m = \boldsymbol{E}\xi(1), |m| < \infty, D\xi(1) < \infty$, $D$ is the variance.

It was shown in Ref. [19] that for $m = k'(0) = 0,\ \rho_\pm(s) \approx \sqrt{2s / D\xi(1)}$, and for $m > 0\ \ (m < 0)$ $\rho_+(s) \approx m^{-1}s\ \ (\rho_-(s) \approx |m|^{-1}s\,),\quad s \to 0$. If $\pm m$<0, then $\rho_\pm(s)_{s\to 0} \to \rho_\pm > 0$.

If a process with positive jumps is monotonically nondecreasing, then Eq. (6) has no negative roots. If a process with negative jumps is monotonically nonincreasing, then equation (6) with $s > 0$ has no positive roots.

**2. 2. Review of the relations of thermodynamics of trajectories**

The partition function (or the MGF) from relations Eqs. (1)-(2), for example, in the thermodynamics of trajectories (Refs. [23-34], below are examples from this area), is equal to:

$$Z_\tau(s) \equiv \sum_K e^{-sK} P_\tau(K) = \sum_{\boldsymbol{X}_\tau} e^{-s\hat{K}[X_\tau]} P(\boldsymbol{X}_\tau) \sim e^{\tau\theta(s)}\,, \qquad \langle\langle K^n \rangle\rangle / \tau = (-1)^n \frac{\partial^n}{\partial s^n}\theta(s)_{|s=0}\,, \tag{7}$$

where $P_\tau(K)$ is the distribution of value K over all trajectories $\mathbf{X}_\tau$ of total time τ, $\mathbf{X}_\tau = (C_0 \to C_{t_1} \to \ldots \to C_{t_K})$, $P(\mathbf{X}_\tau)$ is the probability for observing this trajectory out of all the possible ones of total time τ; $\langle\langle \cdot \rangle\rangle$ indicates cumulant (mean, variance, etc.). The value K defined as the total number of configuration changes in a trajectory. This trajectory has K jumps, with the jump between configurations $C_{t_{i-1}}$ and $C_{t_i}$ occurring at time $t_i$, with $0 \leq t_1 \leq \cdot\ \cdot\ \cdot t_K \leq t$, and no jump between $t_K$ and τ. We within the ensemble of trajectories of total time τ. For x-ensemble Eqs. (8)-(9) is kept fixed is the total number of configuration changes K on trajectories $Y_K = (C_0 \to C_{t_1} \to \ldots \to C_\tau)$, where the number of configuration changes is fixed at K, but the time τ of the final K-th jump fluctuates from trajectory to trajectory. The proportionality $Z_\tau(s)$ of the function $\exp[\tau\theta(s)]$ in Eq. (7) is obtained using the theory of large deviations (LD).

The partition function (7) in statistical physics corresponds to CF (2) or MGF (1). Expressions for SCGFs θ(s) Eqs. (7) refer to the so-called s-ensembles with MGF Eq. (1), Refs. [23-31, 33-34, 37], of the thermodynamics of trajectories in which the process time τ is fixed, and the dynamic activity K, the number of events, changes in the trajectory during time τ is a random variable. In Refs. [10-11], the so-called x-ensembles are considered, in which the values of dynamic activity K are fixed, and the time to reach a fixed value of K, FPT τ of reaching level K is a random variable. The partition function, the corresponding MGF for random time, of reaching a fixed value K of dynamic activity is (in x-ensemble, at x=γ):

$$Z_K(\gamma)=\int_0^\infty d\tau e^{-\gamma\tau}P_K(\tau)\,, \tag{8}$$

where $P_K(\tau)$ is the distribution of total trajectory length τ for fixed activity K. In Ref. [42], the quantity corresponding to $Z_K(\gamma)$, the Laplace transform of the FPT distribution, acts as a non-equilibrium partition function. For large K, the MGF also has a large deviation (LD) form:

$$P_K(\tau)\sim e^{-K\Phi(\tau/K)}\,,\qquad Z_K(\gamma)\sim e^{Kg(\gamma)}\,. \tag{9}$$

In Refs. [42-44], the parameter γ (equal to the parameter x from [10-11]) is conjugate to a random FPT in the distribution.

The total fixed number of configuration changes, i.e., the dynamical activity *K* related with the average dynamical activity per unit time at stationarity $\langle k\rangle$, $K\approx t\langle k\rangle$. When *K*, $\tau$ are large, *K*, $\tau$ >> 1, relations (7), (9) have large deviation form. The function *g* is the functional inverse of $\theta$ and vice versa (Ref. [10]):

$$\theta(s)=g^{-1}(s),\quad g(\gamma)=\theta^{-1}(\gamma),\quad s=g(\gamma),\quad \gamma=\theta(s)\,. \tag{10}$$

Definitions (7) and (8) correspond to definition (1).

### 2. 3. General almost upper semi-continuous processes and risk processes

Let $\xi(t)=\xi_1(t)+\xi_2(t)$ is a general almost upper semi-continuous process with $\int_{-\infty}^{0}\Pi(dx)\le\infty$ , for which the cumulant $\Psi(\alpha)$ is determined by the relations:

$$\Psi(\alpha)=\frac{\lambda_3 i\alpha}{c_3-i\alpha}+ai\alpha+\int_{-\infty}^{0}(e^{i\alpha x}-1)\Pi(dx),\quad a\le 0,\quad c_3\ge 0\,. \tag{11}$$

A process $\xi(t),\ t\ge 0$ is called upper continuous if $P\{\xi(t)-\xi(t-0)\le 0\ \forall\ t\}=1$.

In Refs. [16-22], the processes of achieving positive levels *x*>0 are associated with a cumulant of the form (11), and explicit expressions for the boundary functionals are obtained. For the cumulant (11) with $c_3 m=c_3(a-\tilde{\Pi}(0))+\lambda_3,\ \tilde{\Pi}_-(\alpha)=\int_{-\infty}^{0}e^{i\alpha x}\Pi(x)dx,\ \Pi(x)=\int_{-\infty}^{x}\Pi(dy)$ , $\tilde{\Pi}(0)=\tilde{\Pi}_-(0)<\infty$ , the positive root of equation (6) is equal to $r_s=\rho_+(s)=c_3 p_+(s)<c_3$ ; $p_+(s)=P\{\xi^+(\theta_s)=0\}$ , $\xi^+$ is an extremum of the process *ξ(t)*. At $m=a-\tilde{\Pi}(0)+\lambda_3/c_3>0$, $\rho_+(s)_{s\to 0}\to 0,\ q_+(s)=1-p_+(s)_{s\to 0}\to 1,\ \rho'_+(0)=m^{-1}>0$. At *m*<0, $\rho_+(s)_{s\to 0}\to\rho_+=c_3 p_+>0$, Refs. [16-22].

In Section 3, the cumulant of form (11) is used to establish a connection with the risk processes (12)–(15) and the thermodynamics of trajectories.

The classical risk process has the form Refs. [16-22]:

$$\xi_u(t)=u+ct-\sum_{k=0}^{\nu(t)}\xi_k\,,\quad u>0,\quad c>0\,, \tag{12}$$

where *u* is the initial capital, *c* characterizes the intensity of receipt of insurance premiums (premiums), and $\xi_k$>0 are the values of payments (claims). The process $\xi_u(t)$ is also called the reverse risk process. For classical risk processes with *u*=0, *c*>0, the linear premium function *c(t)=ct* is approximated by a stepwise Poisson process with parameter $\lambda_n$=*cn* (*n*>0) and cumulant:

$$\Psi_n(\alpha)=\lambda_n\left(\frac{n}{n-i\alpha}-1\right)=cn\frac{i\alpha}{n-i\alpha}\Big|_{n\to\infty}\to ci\alpha\,;$$

$S(t)=\xi(t)=\sum_{k\le\nu(t}\xi_k$, $P\{\xi_k>0\}=1$, $E\xi(1)=1/b$, $\xi_k>0$ are the values of payments (claims), $P\{\nu(t)=k\}=(\lambda t)^k e^{-\lambda t}/k!$, $\lambda>0$, is the Poisson distribution. Sometimes it is more:

$$\zeta(t)=u-\xi_u(t)=S(t)-ct,\quad \zeta(0)=0,\quad c>0\,, \tag{13}$$

which is called the claim surplus process.

If in the classical risk process, we replace the deterministic linear function *c(t)=ct* by the stochastic process $S_1(t)=\xi_1(t)=\sum_{k\le\nu_1(t}\eta_k$, $\eta_k>0$, $\eta_0>0$, where $\nu_1(t)$ is a simple Poisson process independent of *v(t)* with intensity $\lambda_1$, then

$$\xi_u(t)=u+S_1(t)-\xi(t),\quad \zeta(t)=\xi(t)-\xi_1(t)\,. \tag{14}$$

The processes $\xi_u(t)$ and $\zeta(t)$ are called reserve and claim surplus risk processes with random premiums, respectively. Herewith:

$$\xi_u(t)=u+\xi_1(t)-S(t),\quad u>0,\quad c(t)=\xi_1(t)=\sum_{k\le\nu_1(t}\xi`_k,\quad S(t)=\sum_{k\le\nu_2(t}\xi_k,\quad P\{\xi_k>0\}=P\{\xi`_k>0\}=1, \tag{15}$$

$\nu_{1,2}(t)$ are simple Poisson processes (independent of each other and independent of $\xi_k$ and $\xi`_k$, $k\ge 0$ ) with intensity parameters $\lambda_{1,2}>0$. These issues are considered in more detail, for example, in Refs. [19-22].

Processes of the form (12)-(15) are used below to determine the characteristics of the processes associated with the SCGF of the MGF for the fluctuations of trajectory observables of the form (16). We are interested in studying fluctuations of observables of the trajectory **X** of the form [35]:

$$A(t)=\sum_{x\ne y}a_{xy}N_{xy}(t)\,, \tag{16}$$

where $a_{xy}$ are arbitrary real numbers with $\sum\left|a_{xy}\right|>0$, and $N_{xy}(t)$ are the elementary fluxes, that is, the number of jumps from *x* to *y* up to time *t* in **X**. For a time-integrated current, the coefficients are antisymmetric, while for counting observables (such as the activity), they are symmetric.

## 3. Risk theory and thermodynamics of trajectories

### 3. 1. The characteristic function of a homogeneous process and its physical interpretation

The characteristic function of a homogeneous process $\xi(t), t\ge 0$ is determined by the relation (if $\xi(0)\ne 0$) (2), where for the function $\Psi(\alpha)$, the cumulants of the process $\xi(t), t\ge 0$, the Lévy-Khintchine form is used, Refs. [12-16]: for a homogeneous process $\xi(t), t\ge 0$, $\xi(0)=0$ with independent increments of the cumulant is, Refs. [12-16]:

$$\Psi(\alpha)=i\alpha\gamma-\frac{1}{2}\alpha^2\sigma^2+\int_{-\infty}^{\infty}(e^{i\alpha x}-1-\frac{i\alpha x}{1+x^2})\Pi(dx)\,,\quad \int_{|x\le 1|}x^2\Pi(dx)<+\infty,\quad \sigma^2\ge 0\,, \tag{17}$$

$\gamma$ has an arbitrary sign. Relation (17) describes infinitely divisible distributions, Refs. [12-16]. In Ref. [45], a physical interpretation of the terms of expression (17) is given, which can be interpreted as the Green's function in the momentum representation (if $\alpha$ is taken as the momentum). The terms $i\alpha\gamma$, $\alpha^2\sigma^2/2$ describe, respectively, the wave motion in the system and the motion of free particles or quasiparticles; $\gamma$ is the speed of sound, $\sigma^2$ is the effective mass. This is consistent with the general definition of $\gamma$ as a real variable and $\sigma^2$ as a non-negative constant. The terms under the integral in (17) describe the interactions of waves with waves, particles with particles, and waves with particles. The function $\Pi(dx)$, which is a measure of the magnitude and number of jumps in the process, characterizes these interactions. In general, the measure $\Pi(dx)$ depends on time.

The Lévy-Khintchine exponent for a shifted compound Poisson distribution with shift $c$ is given by $\Psi(\theta)=\lambda\int_R\left(1-e^{i\theta x}\right)F(dx)-ic\theta$. If further, the shift is chosen to center the compound Poisson distribution, then $c=\lambda\int_R xF(dx)$ and then $\Psi(\theta)=\int_R\left(1-e^{i\theta x}+i\theta x\right)\lambda F(dx)$. We assume the possibility of an integral representation and a transition from the Markov chain to a continuous process associated with a high density of state space.

Many important results depend on the properties of the process (its discontinuity, continuity, etc.). Infinitely divisible distributions and their limits coincide with the limits of sequences of compound Poisson distributions, which corresponds to the representation of Green's function as a series in Green's functions of noninteracting particles. The representation of the cumulant of the characteristic function $\Psi(\alpha)$ in the form (17) is unique, that is, the correspondence between $\Psi(\alpha)$ and the set $\gamma$, $\sigma^2$, $\Pi(dx)$ is one-to-one.

**3. 2. FPT in the thermodynamics of trajectories**

The time value $\tau(x)$ of the first achievement by the trajectory of a random process $Y(t)$ of a certain given level $x$, Refs. [7-11] is studied not only in the theory of random processes, Refs. [12–15] but also, for example, in the thermodynamics of trajectories, Refs. [23–34]. This article uses examples of processes for the fluctuations of trajectory observables. In Ref. [37], the lower bound for the scaled cumulant generating function θ(s) (7) is written in the form:

$$\theta(s)\ge\theta_*(s)=<k>(e^{-s<a>/<k>}-1), \tag{18}$$

where the moment cumulant generating function at long times has large deviation forms (7), $<k>=\sum_{xy}\rho_x W_{xy}$ is the average dynamical activity (per unit time), $\rho_x$ is the stationary distribution, $W_{xy}$ the transition rate from configuration $x$ to $y$, $<a>=\sum_{xy}\alpha_{xy}\rho_x W_{xy}$, $a\approx A/t$.

The trajectory observables are defined in Ref. [37] in terms of the jumps in a trajectory:

$$A(\omega)=\sum_{xy}\alpha_{xy}Q_{xy}(\omega), \tag{19}$$

how in (16), where $Q_{xy}(\omega)$ is the number of jumps from $x$ to $y$ in trajectory $\omega$; $\alpha_{xy}\ge 0$ (unlike (16), where $a_{xy}$ are arbitrary real numbers). This means that $A(\omega)$ is non-negative and nondecreasing with time. The total number of jumps or dynamical activity is $K(\omega)=\sum_{xy}Q_{xy}(\omega)$ ($Q_{xy}(\omega)$ corresponds to the value $N_{xy}(t)$ in (16)).

Inverting $\theta_*(s)$ in Eq (18) provides a lower-scaled cumulant generating function $g(\mu)$ (9)-(10):

$$g(\mu)\ge g_*(\mu)=-\frac{<k>}{<a>}\ln(1+\frac{\mu}{<k>}). \tag{20}$$

The Laplace transformed $FPT$ distribution (9) has a large deviation form $Z_A(\mu)=\hat{F}_x(\mu|A)=\int_0^\infty d\tau e^{-\mu\tau}F_x(\tau|A)\approx e^{Ag(\mu)}$, where $F_x(\tau|A)$ is the $FPT$ distribution, Ref. [37].

In the theory of random processes (e.g. Refs. [16-22]) relations similar to Eqs. (8), (9) are used, but they are supplemented by the condition of finiteness of FPT. From the set of existing boundary functionals of the random process, Refs. [5, 18-22] (functionals from selective trajectories associated with the exit of these trajectories beyond certain boundaries), we consider the moment of the first exit for the level x>0 with MGF of the form:

$$T(s,x)=\boldsymbol{E}[e^{-s\tau^+(x)},\tau^+(x)<\infty]=P\{\xi^+(\theta_s)>x\},\quad \tau^+(x)=\inf\{t>0:\xi(t)>x\}. \tag{21}$$

In [5], it is shown that the finiteness condition of FPT can significantly affect the calculated behavioral features of FPT. In distribution $F_x(\tau|A)$, a conditional distribution is also used, but under the condition that the observable A (for example, the dynamic activity K) is given.

From (9), (20), we obtain:

$$\boldsymbol{E}\tau_{\gamma} = -\frac{\partial \ln Z_K(\gamma)}{\partial \gamma} = -< K > \frac{\partial g(\gamma)}{\partial \gamma} = \frac{< K >}{< a > (1+\gamma / < k >)} . \tag{22}$$

**3. 3. Comparison of the results of the risk theory and thermodynamics of trajectories**

At in (17) $i\alpha\gamma - \sigma^2\alpha^2/2 - \int_{-\infty}^{\infty} \frac{i\alpha x}{1+x^2}\Pi(dx) = 0$, $\int_{-\infty}^{0}(e^{i\alpha x}-1)\Pi(dx) = 0$, the cumulant (17) takes the form:

$$\Psi(\alpha) = \int_{0}^{\infty}(e^{i\alpha x}-1)\Pi(dx) . \tag{23}$$

Let us compare the cumulant (23) (for *is*=*α*) with expression (18). We use the definition MGF (21) for $\tau^+(x) = \inf\{t>0 : \xi(t) > x\}$ and the expression for the MGF, Refs. [16, 19]:

$$T(s,x) = \boldsymbol{E}[e^{-s\tau^+(x)}, \tau^+(x) < \infty] = P\{\xi^+(\theta_s) > x\} = e^{-\rho_+(s)x}, \; x \geq 0 . \tag{24}$$

The MGF $T(s,x)$ is determined by the relation, Refs. [16, 19]:

$$T(s,x) = e^{-\rho_+(s)x}, \quad \rho_+(s) = k^{-1}(s), \quad x \geq 0 , \tag{25}$$

$$\boldsymbol{E}\tau^+(x) = -\frac{\partial \boldsymbol{E}[e^{-s\tau^+(x)-u\gamma^+(x)}, \tau^+(x) < \infty]}{\partial s}\Big|_{u=0,s=0} = \frac{\partial \rho_+(s)}{\partial s}\Big|_{s=0} x e^{-\rho_+ x}, \; x > 0 .$$

The index $\rho_+(s)$ is the positive root of the cumulant equation (6) of the form $k(r) = \Psi(-ir) = s$.

In expression (23) we substitute:

$$\Pi(dx) = < k > \delta(x - < a > / < k >)dx . \tag{26}$$

Then for *iα*=*r* (as in (6)) we obtain:

$$\Psi(\alpha)\big|_{i\alpha=r} = k(r) = < k > (e^{r<a>/<k>} - 1) . \tag{27}$$

Substitution *r*=-*s* in (27) leads to the equality of expressions (27) and (18) and to the relation:

$$\theta(s)=k(r=-s). \tag{28}$$

If we put in the Lundberg equation (6) *s*=*γ*, then we will find from the equation *k(r)*=*γ*= $< k > (e^{r<a>/<k>} - 1)$, assuming $r=r_s=\rho_+(\gamma)$ that:

$$\rho_+(\gamma) = \frac{< k >}{< a >}\ln(1 + \frac{\gamma}{< k >}) . \tag{29}$$

Comparing expressions (29) and (20) for *μ*=*γ*, we obtain:

$$\rho_+(\gamma) = -g(\gamma) . \tag{30}$$

Note that similar relations between the solutions of the analog of the Lundberg equation (6) and the SCGF of FPT are obtained in Ref. [41]. Many quantities and relations from Ref. [41], which also appear in other papers on statistical physics, receive a probabilistic interpretation. The derivation of such quantities, for example, the splitting probability $p_-$ and the generating functions $g_+$ and $g_-$ Ref. [41] are determined by both the Perron root of the tilted matrix and its corresponding right eigenvector, can be described probabilistic using approaches Refs. [12–22].

The correspondence between the expressions (23)-(30) written above and the expression (10) has been established. Thus, the expression $s = g(\gamma)$ corresponds to expressions (28)-(29), since $s = g(\gamma) = -\rho_+(\gamma) = -r_s = s$ , (28)-(29). The expression $\gamma = \theta(s)$ from (10) corresponds to equation (6) with $s = \gamma$ and relation (28); $\gamma = k(r = -s) = \theta(s)$ . The boundary value $K$ from expressions (7)-(9), (22) corresponds to the value $x$ from Eq. (31).

Thus, the solutions to the problem of determining the random variable *FPT* $\tau^+(x)$ obtained in the theory of random processes (23)-(30) and in the thermodynamics of trajectories (7)-(10), (20), (22) coincide under conditions (25)-(30). This happens because the expression (25) of the theory of random processes coincides with the relations (9)-(10) of the thermodynamics of trajectories.

In Refs. [16-22] more complex relations than (25) for MGF $\tau^+$ for are considered. So, in Ref. [19] it is shown that for an almost upper semi-continuous process $\xi(t)$ with SCGF (11) the ratio is established for the pair $\{\tau^+(x), \gamma^+(x)\}$:

$$\boldsymbol{E}[e^{-s\tau^+(x)-u\gamma^+(x)}, \tau^+(x) < \infty] = \boldsymbol{E}[e^{-u\gamma^+(x)}, \xi^+(\theta_s) > x] = \boldsymbol{E}e^{-u\gamma^+(x)} P\{\xi^+(\theta_s) > x\} = \frac{c_1}{c_1 + u} q_+(s) e^{-\rho_+(s)x}, \; x > 0, \quad (31)$$

where $\gamma^+(x) = \xi(\tau^+(x)) - x$ is the first overjump over $x$>0. For MGF (21) in Ref. [19] an integral equation with the solution is derived (as from (31) at u=0):

$$T(s,x) = q_+(s) e^{-\rho_+(s)x}, \quad q_+(s) = 1 - p_+(s), \; \rho_+(s) = c_1 p_+(s) < c_1, \; p_+(s) = P\{\xi^+(\theta_s) = 0\};$$

for $m$<0, $\rho_{+|s\to 0} \to \rho_+ = c_1 p_+ > 0$, where $\rho_+$ is the positive root $r_{s0}$ of the equation $k(r_{s0})$=0, $c_1 \geq 0$, $\xi^+(\theta_s) = \sup_{0 \leq u \leq \theta_s} \xi(u)$ is the extremum of $\xi(u)$ . Note that relation (28) does not hold for expression (11), i.e. it is not a universal relation.

If in (23) at $i\alpha = r$ we specify $\Pi(dx)$ not in the form of (26), but in the form of an exponential distribution that takes into account fluctuations in the magnitude of the jumps:

$$\Pi(dx) = \langle k \rangle m_{3+} e^{-m_{3+}x} dx, \quad m_{3+} = \langle k \rangle / \langle a \rangle, \quad (32)$$

then at $r < m_{3+}$:

$$k(r) = \langle k \rangle r / (m_{3+} - r). \quad (33)$$

From the equation $k(r) = \gamma$ (6) for $r = r_s = \rho_+(\gamma)$ we get:

$$\rho_+(\gamma) = \gamma m_{3+} / (\langle k \rangle + \gamma), \quad \partial\rho_+(\gamma)/\partial\gamma = m_{3+} \langle k \rangle / (\langle k \rangle + \gamma)^2, \quad (34)$$

which also corresponds to (10) for $k(r=-s)=\theta(s)$.

For $m$>0 ($m = \langle a \rangle > 0$), $\rho_{+|s\to 0} \to 0, \quad \rho_+ s^{-1}{}_{|s\to 0} \to 1/m$.

If for relations (25), (29) $\dfrac{\partial\rho_+(\gamma)}{\partial\gamma} = \dfrac{1}{\langle k \rangle} \dfrac{1}{(1 + \gamma/\langle k \rangle)}$, and from (22):

$$\boldsymbol{E}\tau^+(x) = \frac{1}{\langle a \rangle} \frac{x}{1 + \gamma/\langle k \rangle}\Big|_{\gamma=0} = \frac{x}{\langle a \rangle}, \quad (35)$$

then for distribution (23), (32) with cumulant (33) and $\rho_+$ type (34):

$$\boldsymbol{E}\tau^+(x) = \frac{x m_{3+} \langle k \rangle}{(\langle k \rangle + \gamma)^2}\Big|_{\gamma=0} = \frac{x}{\langle a \rangle}. \quad (36)$$

The expressions (35) and (36) for $\boldsymbol{E}\tau^+(x)$ coincide at γ=0. But, if we do not assume γ=0, but use relations of the form of expressions obtained from (8), $\ln Z_K(\gamma) = Kg(\gamma)$ , $\boldsymbol{E}\tau^+(x) = -\partial \ln Z_K(\gamma)/\partial\gamma = -K\partial g(\gamma)/\partial\gamma$, and consider the parameter γ, as in [6, 42-44], as a field conjugate to FPT, then for relations (8)-(9) and (24)-(25), correspondence of the form is established:

$$T(s=\gamma, x=K) \to Z_K(\gamma), \quad -\frac{\partial \ln T(s=\gamma, x=K)}{\partial \gamma} = \boldsymbol{E}\tau^{+}{}_{\gamma}(x) = \frac{x}{<a>}\frac{1}{(1+\gamma/<k>)}. \qquad (37)$$

And for expressions (32)-(34), (36):

$$\boldsymbol{E}\tau^{+}{}_{\gamma}(x) = -\frac{\partial \ln T(s=\gamma, x=K)}{\partial \gamma} = \frac{x}{<a>}\frac{1}{(1+\gamma/<k>)^2}. \qquad (38)$$

The functions $\boldsymbol{E}\tau^{+}{}_{\gamma}(x)$ depending on the parameter $\gamma$, (37), and (38), are different from each other.

Thus, a correspondence of the form is established between the quantities of the theory of random processes and the thermodynamics of trajectories

$$K \to x, \quad g(\gamma) \to -\rho_{+}(\gamma), \quad \boldsymbol{E}[e^{-\gamma\tau^{+}(x)}, \tau^{+}(x)<\infty] \to Z_K(\gamma) \sim e^{Kg(\gamma)}. \qquad (39)$$

For expression (31) for $x>0$ from last relation in (39) we obtain

$$g(\gamma) = -\rho_{+}(\gamma) + \ln[1-\rho_{+}(\gamma)/c_1]/x. \qquad (40)$$

It can be seen from expression (40) that the second relation in (39) is not satisfied, and the term $\ln[1-\rho_{+}(\gamma)/c_1]/x$ is added to $-\rho_{+}(\gamma)$. Relations (39) are valid for the cumulant of the form (20), (23), and for the MGF of the form (25). But for large values of x, the second term on the right-hand side of expression (40) is much smaller than the first term, and relation (39) is satisfied.

Let us note the differences between the approaches of the theory of random processes and the thermodynamics of trajectories. So, for the $x$-ensemble, the fixed value of the value of $K$, which is reached in a random time *FPT*, is positive in the thermodynamics of trajectories, and in the theory of random processes, the value of the limit $x$ achieved by a random process can be both positive and negative.

### 3.4. Similarities and differences between the approaches of the theory of random processes and the thermodynamics of trajectories

The average value of the *FPT* at the level $x>0$ obtained from expression (31) ($m<0$) is equal to:

$$\boldsymbol{E}\tau^{+}(x) = -\frac{\partial \boldsymbol{E}[e^{-s\tau^{+}(x)-u\gamma^{+}(x)}, \tau^{+}(x)<\infty]}{\partial s}\Big|_{u=0,s=0} = \frac{\partial \rho_{+}(s)}{\partial s}\Big|_{s=0}[\frac{1}{c_1} + x(1-\frac{\rho_{+}}{c_1})]e^{-\rho_{+}x}, \; x>0. \qquad (41)$$

For values x>0 (corresponding to $K>0$), we rewrite expression (31) by u=0 in the form:

$$\boldsymbol{E}[e^{-\gamma\tau^{+}(x)}, \tau^{+}(x)<\infty] = (1-\rho_{+}(\gamma)/c_1)e^{-\rho_{+}(\gamma)x}, \; x>0, \qquad (42)$$

where the parameter s is replaced by $\gamma$.

In expression (41) the mean FPT of the factor x from (25) is replaced by the factor $[1/c_1 + x(1-\rho_{+}/c_1)]$. If $1/c_1 = 0$, then this factor is x.

From (31) we get

$$\boldsymbol{E}[e^{-s\tau^{+}(x)-u\gamma^{+}(x)}, \tau^{+}(x)<\infty]\Big|_{u=0,s=0} = \boldsymbol{E}[\tau^{+}(x)<\infty] = (1-\frac{\rho_{+}}{c_1})e^{-\rho_{+}x}, \; x>0. \qquad (43)$$

It is clear that in this case there is a non-zero probability of an infinite lifetime.

The thermodynamics of trajectories with observable $K$ does not take into account the processes of reaching negative levels and does not take into account the values $1/c_1$ that change the form of the characteristic function (31) and the average value of the *FPT* (41). For $m<0$, $\rho_{+} = \rho_{+}(s=0) \neq 0$, but $g(\gamma=0)=0$. In the thermodynamics of trajectories, the case *m*>0 is considered, when $\rho_+=0$, and the case *m*<0 is not considered. The results obtained in Sections 3.3 and 3.4 for *m*<0 is not written using the thermodynamics of trajectories.

For values *x*>0 (corresponding to $\boldsymbol{E}K>0$), we rewrite expression (31) in the form (42).

Let us compare, as above, the MGF (8) with the Laplace transform of the FPT distribution with the characteristic function FPT of the form (42). We define the mean values of FPT in the form corresponding to the relations used in statistical physics for the logarithm of the statistical sum (MGF), as in (37), (38):

$$\boldsymbol{E}\tau^+{}_\gamma(x)=-\frac{\partial\ln\boldsymbol{E}[e^{-\gamma\tau^+(x)},\,\tau^+(x)<\infty]}{\partial\gamma}=\frac{\partial\rho_+(\gamma)}{\partial\gamma}[x+\frac{1}{c_1(1-\rho_+(\gamma)/c_1)}],\;\; x>0. \tag{44}$$

To determine the explicit dependence on the parameter *γ* in (44), it is necessary to solve equation (6) and determine the explicit form of the function $\rho_+(\gamma)$. However, for large values of *x*=*K*, corresponding to large deviations, the second term in (44) can become much smaller than the first, and we obtain expressions (25), (45).

To determine the explicit form of the function $\rho_+(\gamma)$, one can use the approximations of Cramer-Lundberg, Rényi, De Wilder, diffusion, and exponential approximations, Refs. [16-22].

Assuming in (43)-(44) $1/c_1=0$, we obtain:

$$\boldsymbol{E}\tau^+{}_\gamma(x)=x\frac{\partial\rho_+(\gamma)}{\partial\gamma}. \tag{45}$$

which corresponds to Eq. (25) for *x*>0.

This coincidence is because the Lundberg equation (6) and condition (39) correspond to transformation (10). This match seems to hold true in many situations. But in most cases, it is not possible to find an explicit form of the solution. In these situations, one can use asymptotic approximations, such as the Cramer-Lundberg approximation. It seems promising to use expressions similar to (44), when the average value of *FPT* is sought not by differentiating the characteristic function for *FPT* and setting the differentiation argument to zero, but by differentiating the logarithm of the characteristic function for *FPT*, as the logarithm of the partition function (8).

In Ref. [38], the correspondence (39) between the characteristic function for the FPT (42) and the statistical sum (8) is confirmed by calculations. In this case, the parameter γ plays the role of a physical field conjugate to the random thermodynamic parameter of the FPT, Refs. [6, 42-44]. The use of the results of the theory of random processes with the involvement of physical relationships similar to the use of the statistical sum, the Legendre transformation, etc., seems promising.

In the thermodynamics of trajectories, Refs. [24, 29, 33], the function θ(s) is considered as the (negative) dynamic free energy per unit time. In case (16), negative values of random variables are also possible. Then the connection between the characteristic function and the Laplace transform becomes more complicated. If there is a distribution density *p(x)* of a random variable *ξ*, *dF(x)=p(x)dx*, where $F(x)=P\{\xi<x\}$ is the distribution function of a random variable *ξ*, then the characteristic function of a random variable *ξ* is equal to

$$\boldsymbol{E}(e^{i\alpha\xi})=\int_{-\infty}^{\infty}e^{i\alpha x}p(x)dx\Big|_{i\alpha=-s}=\int_0^{\infty}e^{-sx}p(x)dx+\int_{-\infty}^{0}e^{-sx}p(x)dx\,, \tag{46}$$

i.e., the characteristic function of the random variable *ξ* is equal to the sum of the Laplace transform of the distribution density $p(x\geq 0)$ of the random variable *ξ>0* with the change of the argument *iα=-s* and the Laplace transform of the distribution density *p(x<0)* of the random variable *ξ<0*.

The second term on the right side of Eq. (46) has no correspondence in the thermodynamics of trajectories. But in the thermodynamics of trajectories, in the *s*-ensemble the random variable *K*>0, and in the *x*-ensemble the random time *τ*>0. Therefore, the second term on the right-hand side of Eq. (46) is insignificant in the *s*-ensemble and *x*-ensemble.

The expression for the average FTP at x>0, obtained from (44) at γ=0, is related to the expression from equation (43) as follows: $\boldsymbol{E}\tau^{+}{}_{\gamma=0}(x)=\boldsymbol{E}\tau^{+}(x)/\boldsymbol{E}\{\tau^{+}(x)<\infty\}$. In Refs. [43-45] it is shown that the quantity $\boldsymbol{E}\tau^{+}{}_{\gamma=0}(x)$ represents the equilibrium, unperturbed average value of the FTP. There are always flows, entropy production, and other effects that are expressed through the parameter γ and reflected in expressions (42), (44).

Relations similar to expression (31) for *x*<0 are written using expression (47), Ref. [19]:

$$\boldsymbol{E}[e^{-s\tau^{-}(x)+u\gamma^{-}(x)},\tau^{-}(x)<\infty]=\boldsymbol{E}[e^{u\gamma^{-}(x)},\xi^{-}(\theta_s)<x]=\boldsymbol{E}e^{u\gamma^{-}(x)}P\{\xi^{-}(\theta_s)<x\}=\frac{c_1}{c_1+u}q_{-}(s)e^{\rho_{-}(s)x},\ x<0, \quad (47)$$

rewritten as

$$\boldsymbol{E}[e^{-\gamma\tau^{-}(x)},\tau^{-}(x)<\infty]=(1-\rho_{-}(\gamma)/b)e^{\rho_{-}(\gamma)x},\ x<0,\quad q_{-}(s)=(1-\rho_{-}(s)/b),$$

$$\boldsymbol{E}\tau^{-}{}_{\gamma}(x)=-\frac{\partial\ln\boldsymbol{E}[e^{-\gamma\tau^{-}(x)},\tau^{-}(x)<\infty]}{\partial\gamma}=\frac{\partial\rho_{-}(\gamma)}{\partial\gamma}[-x+\frac{1}{b(1-\rho_{-}(\gamma)/b)}],\ x<0.$$

$r(s)=r_s=-\rho_{-}(s),\quad r_s(s=0)=0$ is a negative root of the Lundberg equation (6), $r_s=-\rho_{-}(s)<0,\quad m<0,\quad \rho_{-}(s)_{s\to0}\to0,\quad \rho'_{-}(0)=1/|m|,\quad q_{-}(s)=1-p_{-}(s)$.

If *m*>0, then $\rho_{-}\Big|_{s\to0}\to\rho_{-}=bp_{-}\geq0$, $p_{-}(s)=P\{\xi^{-}(\theta_s)=0\}$, $b>0$.

## 4. Estimates of dynamic activity obtained from the relationship between risk theory and thermodynamics of trajectories

Let us illustrate the possibility of applying the above-mentioned connections using the example of dynamic activity K of the form (7), (48) in thermodynamics of trajectories. In Ref. [5], a connection between the risk theory and the thermodynamics of trajectories was noted. However, the type of SCGF of the risk theory, given in Ref. [5] in general form, was not specified. The process corresponding to this SCGF was also not defined. The purpose of this section is to eliminate this shortcoming and to demonstrate some possibilities of the connection obtained.

SCGF θ(s) (7) in the thermodynamics of trajectories, Refs. [23-34] characterizes, as a rule, the dynamic activity K, describing the number of jumps of some process on its trajectory. Follow Ref. [37], we focus on trajectory observables defined in terms of the jumps in a trajectory (19).

An important example of a counting observable is the total number of jumps or dynamical activity Refs. [23-34], when in (19) $\alpha_{xy}$=1:

$$K(\omega)\ =\sum\nolimits_{xy}Q_{xy}(\omega). \quad (48)$$

SCGF of risk process (13)-(15) in [5] is written as:

$$k(r)=t^{-1}\ln\boldsymbol{E}e^{r\zeta(t)}=-cr+\lambda(\boldsymbol{E}e^{r\zeta_1}-1), \quad (49)$$

In order for the process (13)-(14) with k(r) of the form (49) to be equal to the value K(t) (48), we must put in (13)-(14), (49) c=0, $\xi_k=1$. Then we obtain from (49):

$$dF(x)/dx=\delta(x-1),\quad k(r)=\lambda(e^{r}-1). \quad (50)$$

In Refs. [23-34] the expression (7) is defined. Here, *K* represents the dynamical activity, $\tau$ is the total time, $Z_\tau(s)$ is the *MGF* of *K(t)*, and θ(s) is the SCGF of *K*.

In Ref. [37] an expression for θ(s) of the form (18) was obtained. Taking into account Eq. (28), for $\alpha_{xy}$=1, $\langle a\rangle=\langle k\rangle$, λ=$\langle k\rangle$ the relations (50) and (18) coincide.

Substituting expression (50) into Lundberg equation (6) and solving it, we obtain:

$$\rho_+(s)=\ln(1+s/\langle k\rangle)\geq 0\,,\quad m=\partial k(r)/\partial r_{|r=0}=\langle k\rangle>0. \qquad (51$$

Relation $\rho_+(s)=-g(s)$ (30), also obtained in [38], is satisfied, where g(s) (9) is SCGF of the MGF (8). The dependence $\rho_+(s)$ on s of the form (51) at $\langle k\rangle$=1/1.5 0.66… is shown in Fig. 1

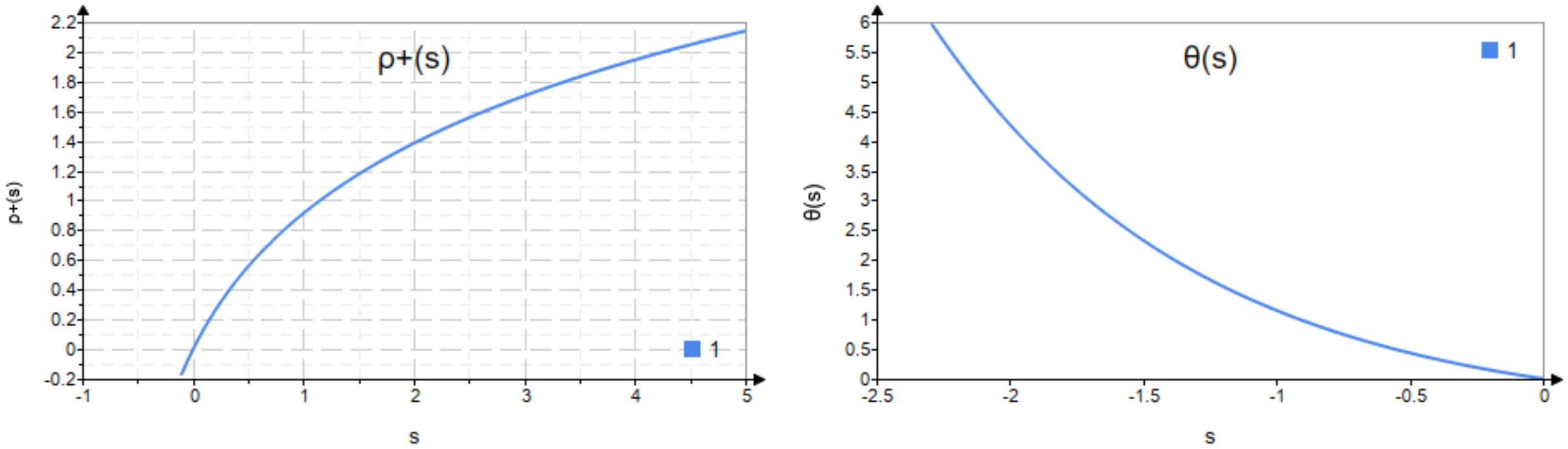

Fig.1. Behavior of the positive root $\rho_+(s)$ (51) of the Lundberg equation (6).

Fig.2. Behavior of the analog of the dynamic free energy *θ(s)* per unit time from *s* in the active phase [28, 33-34]. Fig. 2 can be obtained from Fig. 1 by rotating it to the left and using the resulting expressions $\rho_+(s_{tr})=-s$, $s_{tr}=\theta(s)$.

Since $k(r=-s)=\theta(s)$ (28), then at $r=\rho_+(s_{tr})=-s$, $k(\rho_+(s_{tr})=-s)=\theta(s)$, where we have designated the parameter $s=s_{tr}$ from the risk theory. Using equation (6), $s_{tr}=k(r)$, we obtain that $s_{tr}=k(r=\rho_+(s_{tr})=-s)=\theta(s)$. Thus, the quantities in Fig. 1 take the form: $\rho_+(s_{tr})=-s$, $s_{tr}=\theta(s)$. After such a replacement in Fig. 1 and turning the figure to the left, we obtain Fig. 2, known in thermodynamics of trajectories [23-34] as the behavior of the analog of the dynamic free energy θ(s) per unit time from s in the active phase [28, 33-34]. The quantity g(s) from Eq. (9) can also be plotted along the abscissa axis s, since (10) $g(s)=\theta^{-1}(s)=-\ln(1+s/\langle k\rangle)=-\rho_+(s)=s$.

The equality $\rho_+(s)=-g(s)$ allows us to replace the relationship of thermodynamics of trajectories of the form $\langle\tau_\gamma\rangle=-K\partial g(\gamma)/\partial\gamma$ with the same relationship, but containing the positive root of equation (6): $\langle\tau_\gamma\rangle=K\partial\rho_+(\gamma)/\partial\gamma$. The same applies to the relationship $k(r=-s)=\theta(s)$, when instead of the expression of thermodynamics of trajectories we can use the expression $K=-\tau\partial k(-s)/\partial s$.

What consequences can be obtained from the noted connection between the quantities of risk theory and the thermodynamics of trajectories? Let us indicate which boundary functionals can be applied to the dynamic activity K (48). This quantity is notdecreasing. Therefore, such functionals associated with the possibility of process decrease, such as return moments, time spent above the level, etc., are not applicable to the process K(t). The boundary functionals of the time of first reaching the level, extreme values, the moments of the extremum, and others are applicable.

In order to find the time of the first achievement $\tau^+$ of a positive level u for arbitrary distribution functions of the process jumps, we use the relations obtained for $T(s,u)=\boldsymbol{E}[e^{-s\tau^+(u)},\tau^+(u)<\infty]$ the form (25). The quantity (25) plays the role of the statistical sum (8)-(9). Substituting (9) at s=γ into (25) yields a relation of the form:

$$\ln T(s,u)=-u\rho_+(s)=Kg(s)\,. \tag{52}$$

Since $g(s)=-\rho_+(s)$ then K=u. From (51), (52) we obtain Eq. (37). In this case $\langle k\rangle=K/\langle\tau\rangle$, and $K/\langle k\rangle=\langle\tau\rangle$ at K=u. Expression (37) in the general case depends on the parameter s, which can be considered as a certain field conjugate to the random variable τ. The expression for $\langle\tau_\gamma\rangle=-K\partial g(\gamma)/\partial\gamma$, obtained from $\rho_+(s)=\ln(1+s/\langle k\rangle)$, and $\rho_+(s)=-g(s)$, is equal to (37), $\langle\tau(s)\rangle=-K\partial g(s)/\partial s=K/(s+\langle k\rangle)$.

Expressions for $\langle\tau(s)\rangle$ were obtained in Ref. [5], where the relationships from risk theory Refs. [16–22] were used, and in Ref. [6], where the dependencies on the relative change in entropy were obtained. The dependence $\langle\tau(s)\rangle$ on s of the form (37) $\langle k\rangle=1$ for and K=50 is shown in Fig. 3a. Fig. 3b shows the dependence $\langle\tau(K)\rangle$ on K for s=0. The value of K represents the level of achievement of the FPT process with the cumulant g(s).

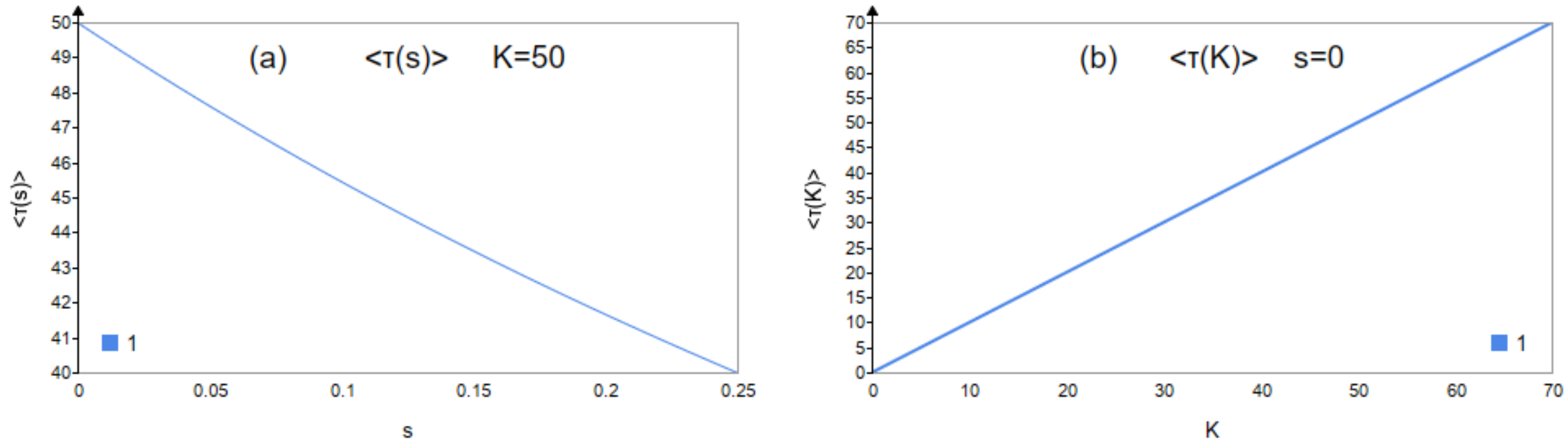

Fig. 3a. Dependence $\langle\tau(s)\rangle$ on s in the range s=0,…,0.25 at K=50 and $\langle k\rangle=1$.

Fig. 3 b. Dependence $\langle\tau(K)\rangle$ on K in the range K=0,…,70 at s=0 and $\langle k\rangle=1$.

In Ref. [42] it is shown that s=0 only in the equilibrium state, where s is the thermodynamic parameter conjugate to FPT (in the notation of Ref. [42] s=γ). Another difference: in the theory of random processes, the characteristic function is differentiated, and in statistical physics its analog, the statistical sum, its logarithm, is differentiated.

The first factorization identity, written in [19], has the form (4)-(5). For $\xi(t)=K(t)$, $i\alpha=-r$, and $T(s,u)=\boldsymbol{E}[e^{-s\tau^+(u)},\tau^+(u)<\infty]$ of form (25) in Ref. [19] the formula is proved:

$$\varphi_+(s,-r)=Ee^{-r\xi^+(\theta_s)}=\rho_+(s)/(r+\rho_+(s))\,. \tag{53}$$

$$E[\xi^+(\theta_s),r]=-\partial\ln E[e^{-r\xi^+(\theta_s)}]/\partial r=(\rho_+(s)+r)^{-1}\,. \tag{54}$$

To go from time $\theta_s$ to time t, we must perform the inverse Laplace-Carson transform with respect to s. For large times t and small values of s, when $\rho_+(s)=\ln(1+s/\langle k\rangle)\approx s/\langle k\rangle$, from (53) we have:

$$\boldsymbol{E}e^{-rK^+(t)}\approx e^{-r\langle k\rangle t}\,. \tag{55}$$

Large values of time t correspond to large values of K=<k>t and the use of LD. Then $\boldsymbol{E}[\xi^+(t),r] = -\partial \ln \boldsymbol{E}[e^{-r\xi^+(t)}]/\partial r = \langle k \rangle t = K(t)$. This corresponds to the non-decreasing value of K, when the maximum value of K will be the last value.

From (5) we have:

$$\varphi_-(s,-r) = \boldsymbol{E}e^{-rK^-(\theta_s)} = \frac{s}{s-\langle k \rangle (e^r-1)} \frac{(r+\rho_+(s))}{\rho_+(s)}. \tag{56}$$

For large times t and small values of s, when $\rho_+(s) = \ln(1+s/\langle k \rangle) \approx s/\langle k \rangle$, $\boldsymbol{E}[e^{-rK^-(\theta_s)}] \approx \dfrac{r\langle k \rangle + s}{s-\langle k \rangle (e^r-1)}$. Performing the inverse Laplace-Carson transform on s, we obtain that $\boldsymbol{E}[e^{-rK^-(t)}] \approx e^{\langle k \rangle t(e^r-1)}(1+r/(e^r-1)) - r/(e^r-1)$. The average value is: $-\partial \ln \boldsymbol{E}[e^{-rK^-(t)}]/\partial r = \dfrac{\langle k \rangle te^r(e^r-1+r)+(1-re^r/(e^r-1))(1-e^{-\langle k \rangle t(e^r-1)})}{e^r-1+r-re^{-\langle k \rangle t(e^r-1)}}$. From this, we obtain that when $r \to 0$:

$$\boldsymbol{E}[K^-(t), r=0] = \langle k \rangle t = K(t). \tag{57}$$

From (55), (57), we have: $\langle K^+ \rangle = \langle K^- \rangle = K(t)$. But $K^+(t) = \sup_{0\le u\le t} K(u)$. The maximum is achieved when u=t, and $\langle K^+(t) \rangle = \langle k \rangle t = K(t)$; $K^- = \inf_{0\le u\le t} K(u)$, the minimum is achieved when u=0, and $\langle K^-(t) \rangle$=0.

The expression for the mean value $\xi^+(\theta_s)$ obtained from (53) is ($\xi^+(\theta_s) = K^+(\theta_s)$) (54):

$$\langle \xi^+(\theta_s), r \rangle = -\partial \ln \varphi_+(s,-r)/\partial r = E[(\xi^+(\theta_s) = K^+(\theta_s))\exp[-rK^+(\theta_s)]]/E[\exp[-rK^+(\theta_s)],] = (\rho_+(s)-r)^{-1}. \tag{58}$$

Note that the inverse Laplace-Carson transform of expression (58) does not coincide with the same expression obtained from expression (55). This is due to the fact that in expression (58), the dependence on s is complex; the value of $\theta_s$ is included in three places in the denominator and numerator. The dependences $\langle K^+(\theta_s), r \rangle$ on r of the form (58) (as above, the parameter r can be considered as a certain field conjugate to the random variable $K^+$) and on s of the form (58) are shown in Fig. 4a and 4b.

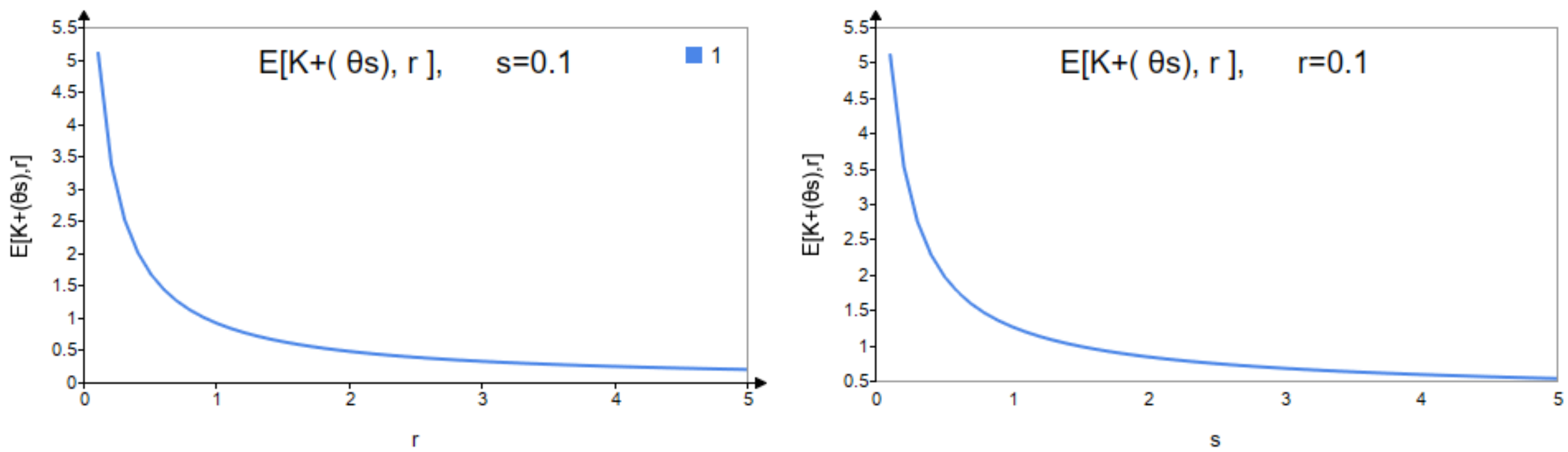

Fig.4. a) Fig. 4a. Dependence $\langle K^+(\theta_s), r \rangle$ (54), (58) on r in the range r=0.1,…,5 for s=0.1 and $\langle k \rangle = 1$.
b) Fig. 4b. Dependence $\langle K^+(\theta_s), r \rangle$ (54), (58) on s in the range s=0.1,…,5 at r=0.1 and $\langle k \rangle = 1$.

The second factorization identity, written in Ref. [19] as a theorem, has the form:

$$\boldsymbol{E}[e^{-s\tau^{+}(\theta`_{\mu})-u\gamma^{+}(\theta`_{\mu})}, \tau^{+}(\theta`_{\mu})<\infty]=\frac{\mu}{\mu-u}\{1-\frac{\varphi_{+}(s,i\mu)}{\varphi_{+}(s,iu)}\}\ \ (s,u,\mu>0)\,. \tag{59}$$

Here $\gamma^{+}(x)=\xi(\tau^{+}(x))-x$ is the first overjump over x>0, $\theta`_{\mu}$ is an exponentially distributed random variable with parameter μ>0, independent of $\theta_{s}$, $\varphi_{+}$ from (5). For MGF of $\gamma^{+}(\theta`_{\mu})$ at s=0, we have from (53):

$$\boldsymbol{E}[e^{-u\gamma^{+}(\theta`_{\mu})}, \tau^{+}(\theta`_{\mu})<\infty]=1\,. \tag{60}$$

If we move from variable $\theta`_{\mu}$ to variable x, then:

$$\boldsymbol{E}[e^{-u\gamma^{+}(t)}, \tau^{+}(x)<\infty]=1\,. \tag{61}$$

From both (60) and (61) we obtain that the average value of the overjump over K>0 for the value of K is equal to zero.

Also, in the case of arbitrary distribution functions of process jumps, relationships are written for process extremes, moments of reaching an extreme, etc. From the non-decrease in dynamic activity K, it follows that the last value of K in time will be the maximum. Therefore, the moment of reaching the maximum value coincides with the time of first reaching the last level.

Note that in Ref. [37] another approach is used to study FPT, which is related to kinetic equations for dynamic quantities. However, in Ref. [37] other boundary functionals besides FPT are not considered.

The results of the thermodynamics of trajectories can also be used in risk theory. As an example, we will cite the Legendre transformations. For $\zeta(t)=K(t)$ in the thermodynamics of trajectories, relations (7) are satisfied, the relation LD, Refs. [37, 47]:

$$P_{t}(K)\approx e^{-t\varphi(K/t)}\,, \tag{62}$$

and also, Legendre-Fenchel transform Ref. [37]:

$$\varphi(k)=-\min_{s}[\theta(s)+ks]\,,\quad \theta(s)=-\min_{k}[\varphi(k)+ks]\,,\quad \langle k(s)\rangle=-\partial\theta(s)/\partial s,\ \ k=K/t\,. \tag{63}$$

From (63) and (50), (7) we have:

$$P_{t}(k=a)\approx e^{-tI(a)}\,,\quad I(a)=\langle k\rangle-a+a\ln(a/\langle k\rangle)\,,\quad \varphi(k)=I(a),\ \ k=a\,, \tag{64}$$

where $P_{t}(k=a)$ from expression (62). For expressions (7), (49)-(50):

$$\boldsymbol{E}e^{rK(t)}=e^{tk(r)}=\sum\nolimits_{K}e^{rK}P_{t}(K)\,. \tag{65}$$

Similar relationships are valid for values (8)-(9). The expression for g(s) corresponds to Example III.2 Ref. [47] In expression (9):

$$P_{K}(\tau)\approx e^{-K\Phi(\tau/K)}\,, \tag{66}$$

where for $\Phi(t)$ the following Legendre-Fenchel transform is valid:

$$\Phi(t)=-\min_{x}[g(x)+tx]\,,\quad g(x)=-\min_{t}[\Phi(t)+tx]\,,\quad \partial g(x)/\partial x=-t\,. \tag{67}$$

From (50), (7), and (67), we have:

$$\Phi(t)=\langle k\rangle t-1-\ln(\langle k\rangle t)\,. \tag{68}$$

The quantity (68) appears in the relation (66) and in the expressions of the theory of risk and thermodynamics of trajectories (8) of the form $\boldsymbol{E}e^{-r\tau^{+}(s)}=\int_{0}^{\infty}e^{-r\tau}P_{K}(\tau)d\tau$.

Another expression can be given for the average time of stay of the process K(t) in the interval (M, K), M<K: $\tau_{MK}=\tau_{K}(1-M/K)$, where $\tau_{K}$ is the mean first-passage time reaching the maximal value K. This expression is substantiated using the results of Ref. [19].

Numerous results for the probability of bankruptcy obtained in risk theory are applicable to estimates of the probability of K reaching a given value. Both in the study of K and other values of thermodynamics of trajectories, for example, different times, factorization identities, various inequalities, for example, Lyapunov, Doob, Lindblad's, Refs. [16-22], and other results of risk theory are possible.

## 5. Conclusion

Above, after equation (30), the similarity with the results of article [41] was noted. But there are also significant differences. Thus, the Lindblad equation (6) in the notation of this article in Ref. [41] is written as k(-r)=-μ. Accordingly, the solutions of this equation, denoted $a_{\pm}$ in Ref. [41], differ from the solutions $\rho_{\pm}$ of equation (6). And although some expressions for $\boldsymbol{E}[e^{-s\tau^{+}(x)}]$ in Eqs. (24)-(25) and in Ref. [41] formally coincide, the difference in $a_{\pm}$ and $\rho_{\pm}$ leads to different results. This is explained by the fact that in Ref. [41] and in Eqs. (24)-(25) there are different conditions in the conditional averages: in Eqs. (24)-(25) the finiteness condition of FPT, and in Ref. [41] the condition of the form $\langle\cdot\rangle_{+}=\langle\cdot|O_T\geqslant\ell_{+}\rangle$ and $\langle\cdot\rangle_{-}=\langle\cdot|O_T\leqslant-\ell_{-}\rangle$ are expectation values conditioned on events terminating at the positive and negative threshold, respectively. In addition, in Eqs. (24)-(25) and in Ref. [41] different domains of change of time-additive observables are defined: in Ref. [41] from $-\infty$ to x>0, and in Eqs. (24)-(25) from 0 to x>0.

Some connections and mutual influences of the theory of random processes and statistical physics, risk theory and thermodynamics of trajectories are also noted in [38], where 11 points of such connections are indicated. Let us note the most important of them.

- The probabilities of the finiteness of *FPT* $E(\tau(x)<\infty)$ are explicitly included in the conditional characteristic functions (e.g., (21), (24), (31), (39), (41)-(44), (47)). These probabilities are usually not taken into account in statistical physics. This circumstance can lead to incorrect results. If, for example, the average value of *FPT* is determined, as is customary in the theory of random processes, using the relation $-\partial E[e^{-s\tau^{+}(u)},\tau^{+}(u)<\infty]/\partial s|_{s=0}$, then the dependence of the average *FPT* on the level of achievement *u*=*x*=*K* will have the form of a decreasing exponential. However, using the relation accepted in statistical physics, of the form $-\partial\ln E[e^{-s\tau^{+}(u)},\tau^{+}(u)<\infty]/\partial s$, leads to a linear increase in the average value of *FPT* depending on *u*=*x*=*K*. This happens because the average value $-\partial E[e^{-s\tau^{+}(u)},\tau^{+}(u)<\infty]/\partial s|_{s=0}$ contains two factors: $E[\tau^{+}(u)]$ and $-\partial E[e^{-s\tau^{+}(u)}]/\partial s|_{s=0}$. A more correct definition in this case is $-\partial\ln E[e^{-s\tau^{+}(u)},\tau^{+}(u)<\infty]/\partial s$.
- In the thermodynamics of trajectories, the function (9) $g(\gamma=0)=0$. But for the case *m*<0, as shown in the article, the function $\rho_{+}=\rho_{+}(\gamma=0)\neq 0$, although $\rho_{+}(\gamma)$ corresponds to the function $-g(\gamma)$ (30) or the relation $g(\gamma)=\ln E[e^{-\gamma\tau^{+}(x)},\tau^{+}(x)<\infty]/x$. This is because situations *m*<0 is not considered in the thermodynamics of trajectories. Risk theory in this regard is more complete than the thermodynamics of trajectories.
- The characteristic function and moments of reaching negative levels *x*<0, as well as cases *m*<0, are obtained. In the thermodynamics of trajectories, the level of achievement is most often compared with the positive value of the dynamic activity *K*, and the moments of reaching negative levels $x$<0, as well as cases $m$<0, are not considered. However, in the general case at the mathematical level, the choice of the observable $A$ is somewhat arbitrary.

- Since a correspondence is assumed between the partition function $Z_K(\gamma)$ (8) and the characteristic function (21), the average values $E\tau$, which in statistical physics are expressed through the statistical sum $-\partial \ln Z_K(\gamma)/\partial\gamma$, are proposed to be expressed through the characteristic function $-\partial \ln E[e^{-\gamma\tau^+(x)}, \tau^+(x)<\infty]/\partial\gamma$, although the definition $-\partial E[e^{-\gamma\tau^+(x)}, \tau^+(x)<\infty]/\partial\gamma\big|_{\gamma=0}$ is used in the theory of random processes.

- After differentiation with respect to $\gamma$, when determining the average values of *FPT*, it is not necessary to assume $\gamma=0$, since this corresponds to the equilibrium case, and in non-equilibrium situations $\gamma\neq 0$. The parameter $\gamma$, conjugated to the thermodynamic parameter *FPT*, Refs. [39-40], plays the role of some external field associated with the change in the entropy of the system. Just as the infinity value of the absolute temperature, at which the parameter $\beta=1/k_BT$ ($T$ is the absolute temperature, $k_B$ is the Boltzmann constant), conjugated to random energy, is equal to zero, is unattainable, so is the absolute equilibrium is unattainable, at which $\gamma=0$. The influence of the parameter $\gamma$ on various nonequilibrium processes is shown in Refs. [5-6, 42-44, 46].

- A combination of approaches from statistical physics and the theory of random processes is proposed, for example, the use of the Legendre transform for cumulants of random processes.

- A correspondence has been established between the characteristic functions of the theory of random processes and the statistical sum $Z_K(\gamma)$ of statistical physics, between the cumulant $\theta(s)$ corresponding to the dynamic free energy per unit of time, and the cumulant of the random process $\Psi(\alpha)$ (2).

- The obtained correspondence between the level $x$ of *FPT* achievement and the activity value $K$ is possible only when $x$>0, since $K$>0.

In Ref. [38] 11 points of similarity and difference between the theory of random processes and thermodynamics of trajectories are noted. There may be others.